\RequirePackage{fix-cm}
\documentclass[smallextended]{svjour3}       % onecolumn (second format)
\smartqed  % flush right qed marks, e.g. at end of proof
\usepackage{latexsym}
\usepackage{amsmath}
\usepackage{amssymb}
\usepackage{ulem}
\usepackage{fancyhdr}
\usepackage{url}
\usepackage{grffile}
\usepackage[vcentering,dvips]{geometry}
\usepackage{sidecap}
\usepackage{graphicx}
\usepackage{placeins}
\usepackage{booktabs,xltabular}
\usepackage{float}

\newcommand{\be}{\begin{equation}}
\newcommand{\ee}{\end{equation}}
\newcommand{\beq}{\begin{eqnarray}}
\newcommand{\eeq}{\end{eqnarray}}

\newcommand{\ba}{\begin{align}}
\newcommand{\ea}{\end{align}}

\begin{document}

\title{Modifications to the signal from a gravitational wave event due to a surrounding shell of matter
%\thanks{Thanks to ...}
}
%\subtitle{Do you have a subtitle?\\ If so, write it here}

%\titlerunning{Short form of title}        % if too long for running head

\author{Monos Naidoo \and
        Nigel~T. Bishop$^*$   \and
        Petrus~J. van der Walt
}

%\authorrunning{Short form of author list} % if too long for running head

\institute{Monos Naidoo \at
             Department of Mathematics, Rhodes University, Grahamstown, 6140, South Africa \\
              \email{monos.naidoo@gmail.com}           %  \\
              \and
Nigel~T. Bishop$^*$ \at
              Department of Mathematics, Rhodes University, Grahamstown, 6140, South Africa \\
              \email{n.bishop@ru.ac.za}           %  \\
             \and
			Petrus~J. van der Walt \at
			Department of Mathematics, Rhodes University, Grahamstown, 6140, South Africa \\
			\email{p.vanderwalt@ru.ac.za}          
 %  \\
}

\date{Received: date / Accepted: date}
% The correct dates will be entered by the editor

%%%%%%%%%%%%%%%%%%%%%%%%%%%%%%%%%%%%%%%%%%%%%%%%%%%%%%%%%%%%%%%%%%%
\maketitle
%%%%%%%%%%%%%%%%%%%%%%%%%%%%%%%%%%%%%%%%%%%%%%%%%%%%%%%%%%%%%%%%%%%
\begin{abstract}
In previous work, we established theoretical results concerning the effect of matter shells surrounding a gravitational wave (GW) source, and we now apply these results to astrophysical scenarios. Firstly, it is shown that GW echoes that are claimed to be present in LIGO data of certain events, could not have been caused by a matter shell. However, it is also shown that there are scenarios in which matter shells could make modifications of order a few percent to a GW signal; these scenarios include binary black hole mergers, binary neutron star mergers, and core collapse supernovae.


% \PACS{PACS code1 \and PACS code2 \and more}
% \subclass{MSC code1 \and MSC code2 \and more}
\end{abstract}
%%%%%%%%%%%%%%%%%%%%%%%%%%%%%%%%%%%%%%%%%%%%%%%%%%%%%%%%%%%%%%%%%%%
\section{Introduction}
\label{intro}
%%%%%%%%%%%%%%%%%%%%%%%%%%%%%%%%%%%%%%%%%%%%%%%%%%%%%%%%%%%%%%%%%%%
In previous work~\cite{Bishop:2019ckc}, we developed a model for the effect of a matter shell around a gravitational wave (GW) source, obtaining an analytic expression for the modifications to the GWs. In this paper, we apply this model to astrophysical GW sources. The modifications found in~\cite{Bishop:2019ckc} included an echo. The idea of a GW echo has received much attention as being a signature of an exotic compact object(ECO), and it has also been investigated in terms of the astrophysical environment.

At the quantum level, Hawking's 
information paradox suggests Planck-scale modifications of black hole horizons 
(firewalls \cite{Almheiri:2012rt}) and other modifications to black hole structure (fuzzballs \cite{Lunin:2001jy}). 
Dark matter particles have been suggested surrounding star-like objects \cite{Giudice:2016zpa}. Other postulates include
stars with interiors consisting of self-repulsive, de Sitter spacetime, surrounded by a shell
of ordinary matter (gravastars \cite{Mazur:2004fk}). Then there are Boson stars, 
which are macroscopic objects made up of scalar fields \cite{Liebling:2012fv}. 
All these objects are compact bodies mimicking black holes, but without a horizon.
One consequence of these horizonless structures is that ingoing gravitational 
waves produced in a merger may reflect multiple times off effective radial potential barriers. The gravitational waves may be, in effect,
trapped between effective radial potential barriers causing the waves to be `bounced' off these barriers several times with
wave packets leaking out to infinity at regular times. These GW signals, `trailing' the main (outward bound) signal  are referred to as GW echoes. 
\cite{Cardoso:2016rao,Cardoso:2016oxy,Cardoso:2017cqb}.
Much further discussion around GW echoes has been within the context of `new physics' \cite{Nakano:2017fvh,Testa:2018bzd,Wang:2018mlp,Bueno:2017hyj,Maselli:2017tfq,Barcelo:2017lnx,Carballo-Rubio:2018jzw}.

Echoes from a massive, thick shell were
considered in \cite{Barausse:2014tra}. This was a generalisation of the approach of \cite{Leung:1999rh} which was based on an infinitely thin shell. As we have shown in our paper ~\cite{Bishop:2019ckc} the case for a thin shell can be generalised to that of a thick shell by considering several concentric thin shells and integrating.  \cite{Barausse:2014tra} showed that the deviation from Schwarzschild ringdown in their astrophysical estimations were relatively small except for a large mass which indicated that for the majority of astrophysical scenarios the effect would be relatively small. However they did note that  considerations of dark matter around black holes (or compact body mergers) would leave some parametric freedom for echoes as well.
\cite{Konoplya2019}, studied both the combination of contributions of  modifications of the Schwarzschild geometry near the surface, and a nonthin shell of matter surrounding the compact body/merger.
They found that a massive shell at a distance could be distinguished from the purely Schwarzschild evolution of perturbations. However, for the situation of  new physics near the surface of a compact object, (a wormhole in their case), the strong echoes of the surface dominate the echoes of the distant shell.   Furthermore, they found that it would take an extraordinarily large mass,  located sufficiently close to the wormhole, to lead to  discernable changes in the main echoes of the surface and that these changes would be relatively small.
The interaction of GWs with matter has also been studied in cosmology~\cite{Goswami17,Baym17}, with the objective of using GW observations to constrain the properties of dark matter.

It has been suggested that GW echoes have already been observed in the LIGO data of the binary black hole merger GW150914~\cite{Abedi:2016hgu,Conklin:2017lwb}, and also from the Binary Neutron Star (BNS) merger GW170917~\cite{Abedi:2018npz}. 
These claims have been contested \cite{Ashton:2016xff,Westerweck:2017hus}, sparking a debate and responses in defense of the claims ~\cite{Abedi:2017isz,Abedi:2018pst} with further substantiations ~\cite{Abedi:2018npz,Abedi:2020sgg}.

The plan of this paper is as follows. In previous work \cite{Bishop:2019ckc}, we showed that a thin spherical dust shell surrounding a GW source, causes the GW to be modified both in magnitude and phase, but without any energy being transferred to or from the dust. That work suggests the possibilty of GW echoes. In Sec.~\ref{s-back} we describe the problem considered in this scenario, the assumptions made and the key results. 

The solution of~\cite{Bishop:2019ckc} is for a monchromatic GW source. A general waveform may be decomposed into a sum of Fourier components, and the technical details are given in Sec.~\ref{s-FT}. The  decomposition is implemented within a Matlab script using the Fast Fourier Transform, and validation results are reported in this section.

Secs.~\ref{s-echoes} and \ref{s-GW-other} are about astrophysics. First in Sec.~\ref{s-echoes}, we investigate whether a matter shell could explain the echoes that may exist in the LIGO data of GW150914 and GW170817. It is found that the shell would need to have such a large mass that it would constitute a black hole. It follows that if GW echoes are confirmed in another GW event, and with a relative magnitude and delay time similar to that of GW150914 and GW170817, then a matter shell would not be a viable explanation, so strengthening the case that an ECO would have been observed.

It will be shown that a key factor in determining the echo properties of a matter shell is the echo delay time, which in the cases above was of order $1$s. If the echo delay time is much smaller, order $1$ms or smaller, then shell properties that are physically acceptable could lead to measurable effects; however, the short delay time would mean that the effect would not appear as an echo in the usual sense, but rather as a modification to the original signal. Examples of such signal modifications are given in Sec.~\ref{s-GW150914} for a matter shell around an event like GW150914; in Sec.~\ref{ss-BNS} for black hole quasinormal mode signal following a binary neutron star merger; and in Sec.~\ref{ss-supernova} for the case of core collapse supernovae (CCSNe).

 Our conclusions are discussed in Sec~\ref{s-conc}. We also provide, in Appendix~\ref{a-Matlab}, a summary of the Matlab scripts used; these scripts are available as Supplementary Material.

%%%%%%%%%%%%%%%%%%%%%%%%%%%%%%%%%%%%%%%%%%%%%
\section{Effect of a shell of matter on a GW}
\label{s-back}
\begin{figure}
	\begin{center}
		\includegraphics[width=0.50\columnwidth,angle=0]{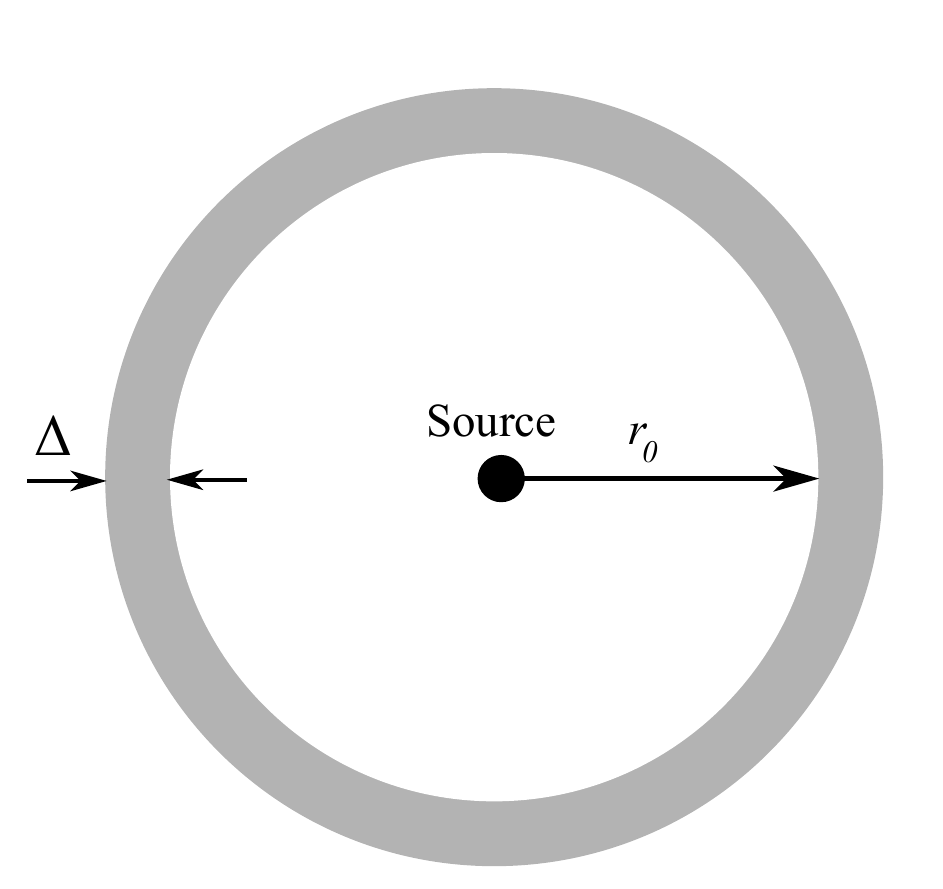}
		\caption{The problem of a GW source which is surrounded by a spherical shell of mass $M_S$ located between $r=r_0$ and $r=r_0+\Delta$, where $r$ is the distance from the source.}
		\label{f-schematic}
	\end{center}
\end{figure}

%%%%%%%%%%%%%%%%%%%%%%%%%%%%%%%%%%%%%%%%%%%%%
In ~\cite{Bishop:2019ckc} we considered the scenario of a thin shell of matter surrounding a gravitational wave source such as a compact binary merger, as shown schematically in Fig.~\ref{f-schematic}. The spacetime around the GW source in ~\cite{Bishop:2019ckc} would be otherwise empty except for the surrounding shell of matter. Confining the investigation to a thin shell does not preclude the case of a thick shell. Results can easily be applied to a series of concentric thin shells and then integrated to give the effect for a thick matter shell. The EOS is taken, as a start, to be that of dust.  The results show that the shell  modifies the outgoing GWs in both phase and magnitude without contradicting previous results about energy transfer.
%%%%%%%%%%%%%%%%%%%%%%%%%%%%%%%%%%%%%%%%%%%%%
The problem is set up within the Bondi-Sachs formalism for the Einstein equations with coordinates based on outgoing null hypersurfaces
~\cite{Bondi62,Sachs62}. The null hypersurfaces, are labelled by the coordinate $x^0=u$, the angular coordinates by $x^A$ ($A=2,3$) and the surface area radial coordinate by $x^1=r$.  The angular coordinates (e.g. spherical polars $(\theta,\phi)$) label the null ray generators of a hypersurface $u=$ constant.  The Bondi-Sachs metric then describes a general spacetime, which may be written as
%%%%%%%%%%%%%%%%%%%%%%%%%%%%%%%%%%%%%%%%%%%%%
\begin{align}
	ds^2  = & -\left(e^{2\beta}\left(1 + \frac{W}{ r}\right)
	- r^2h_{AB}U^AU^B\right)du^2
	- 2e^{2\beta}dudr \nonumber \\
	& - 2r^2 h_{AB}U^Bdudx^A
	+  r^2h_{AB}dx^Adx^B\,,
	\label{eq:bmet}
\end{align}
where, $h^{AB}h_{BC}=\delta^A_C$.
The condition that $r$ is a surface area coordinate implies $\det(h_{AB})=\det(q_{AB})$, where, $q_{AB}$ is a unit sphere metric (e.g. $d\theta^2+\sin^2\theta d\phi^2$).

%%%%%%%%%%%%%%%%%%%%%%%%%%%%%%%%%%%%%%%%%%%%%
%%%%%%%%%%%%%%%%%%%%%%%%%%%%%%%%%%%%%%%%%%%%%
The GW strain far from the source is written $\mathcal{H}_{M0}=r(h_+ +ih_\times)$, where $h_+,ih_\times$ are the usual polarization modes in the TT gauge. Now suppose that, in the absence of the matter shell,
\begin{equation}
	{\mathcal H}_{M0}=\Re(H_{M0} \exp(2i\pi f u))\,{}_2Z_{2,2}\,,
	\label{e-HM0}
\end{equation}
where $f$ is the frequency (assumed to be monochromatic) of the GWs; $H_{M0}$ is a constant determined by the physics of the GW source; and ${}_2Z_{2,2}$ is a spin-weighted spherical harmonic related to the usual ${}_sY_{\ell,m}$ as specified in~\cite{Bishop2016a,Bishop-2005b}. Then, as found in~\cite{Bishop:2019ckc}, the introduction of a spherical shell around the GW source of mass $M_S$, radius $r_0$ and thickness $\Delta$ modifies the wave strain to:
\begin{equation}
	{\mathcal H}=
	\Re\left(H_{M0}\left(
	1+\frac{2M_S}{r_0}+\frac{iM_S}{\pi r_0^2 f}+\frac{i M_Se^{-4i\pi r_0 f}}{4\pi r_0^2f}+{\mathcal O}\left(\frac{M_S\Delta}{r^2_0},\frac{M_S}{r_0^3 f^2} \right)\right)\exp(2i\pi f u)\right)\,{}_2Z_{2,2}\,.
	\label{e-Nshell}
\end{equation}
Each of the terms containing $M_S$ in Eq.~(\ref{e-Nshell}) represents a correction to the wave strain in the absence of the shell~\cite{Bishop:2019ckc}. The first correction, $2M_{S}/r_{0}$, is part of the gravitational red-shift effect, the main consequence of which is a reduction in the frequency; this effect is well-known, and henceforth we will assume that GW waveforms to be considered have allowed for this effect. The second term, $iM_S/(\pi r_0^2 f)$, is out of phase with the leading terms $1+2M_{S}/r_{0}$ and hence represents a phase shift of the GW. This term, to ${\mathcal O}(M_S)$, does not change the magnitude  of  ${\mathcal H}$ and hence has no effect on the energy of the GW. 
The presence of $e^{-4\pi ir_0 f}$ in the third term  describes a change in the magnitude  of  ${\mathcal H}$, as verified in ~\cite{Bishop:2019ckc}. In this context, the modified signal would then be interpreted as an echo of the main signal. The echo varies from the main signal in both magnitude and phase, with the magnitude of the echo  described by
\begin{equation}
	R=\frac{M_S}{4\pi r_0^2 f}
	\label{e-MagEcho}
\end{equation}
relative to the original signal.
%%%%%%%%%%%%%%%%%%%%%%%%%%%%%%%%%%%%%%%%%%%%%
\section{Solution for a burst-like GW source using Fourier transforms}
\label{s-FT}
%%%%%%%%%%%%%%%%%%%%%%%%%%%%%%%%%%%%%%%%%%%%%
The time-delay of the echo is $2r_0$, but the echo's magnitude depends on the wave frequency $f$. 
The GW sources reported to date are not monochromatic but are burst-like. Such a source may be decomposed into its Fourier components and Eq.~(\ref{e-Nshell}) applied to each component, and the echo signal obtained by summing over the transformed components. 
Because the magnitude of the transformation is frequency-dependent, the echo signal will have a form more complicated than simply a time-delay and magnitude change to the original signal. 
This effect is now analyzed.

We replace $\Re(H_{M0}\exp(2i\pi f u))$ in Eq.~(\ref{e-HM0}) by $h(u)$ defined in the interval $u_0\le u\le u_{N-1}$; and then construct a discrete representation of $h(u)$, $h_k=h(u_k)$ ($k=0,\cdots,N-1$), with the $u_k$ on a regular grid, i.e. $u_{k+1}-u_k=\delta$ for $k=0,\cdots,N-2$. Note that if the highest frequency that needs to be resolved is $f_m$, then $N$ should be chosen so that $(u_{N-1}-u_0)/(N-1)< 1/(2f_m)$, i.e. to satisfy the Nyquist condition. The discrete Fourier transform~\cite{Press92} of $\{h_k\}$ is
\begin{equation}
	H_n=\sum_{k=0}^{N-1} h_k \exp\left(\frac{2\pi i k n}{N}\right)\;\;\mbox{with inverse}\;\;
	h_k=\frac 1N \sum_{n=0}^{N-1} H_n \exp\left(\frac{-2\pi i k n}{N}\right)\,.
\end{equation}
Then defining $H_{2,n}$ and $H_{3,n}$ to be coefficients in the transform domain of the second (phase-shift) and third (echo) terms of Eq.~(\ref{e-Nshell}), we have
\begin{align}
	H_{2,n}&=\frac{iM_S H_n}{\pi r_0^2 f_n}\,,\;\;
	H_{3,n}=\frac{iM_S H_n\exp(-4\pi ir_0f_n)}{4\pi r_0^2f_n}\,,\;\;
	n=1,\cdots,\frac N2\,,\nonumber\\
	H_{2,n}&=H_{2,N-n}^*\,,\;\;H_{3,n}=H_{3,N-n}^*\,,\;\; n=\frac N2 +1,\cdots,N-1\,,\nonumber\\
	H_{2,0}&=H_{3,0}=0\,,
\end{align}
where ${}^*$ denotes the complex conjugate, and where we have used the condition that, in the time domain, all quantities are real.
It is being assumed that $N$ is even, and normally $N=2^m$ (with $m$ an integer) for convenience when using the fast Fourier transform; further
\begin{equation}
	f_n=\frac{ n(N-1)}{N(u_{N-1}-u_0)}\,.
\end{equation}
Then $h_{2,k},h_{3,k}$ are found on applying the inverse discrete Fourier transformation.

A Matlab script that implements the calculation of the previous paragraph  is available as Supplementary Material,  
and is described in Appendix~\ref{a-Matlab}. 
The script was checked by applying it to a monochromatic signal $h(u)=\Re(-i\exp(2\pi i f u))$. The errors $e_{2,k}$  in $h_{2,k}$ and $e_{3,k}$ in $h_{3,k}$ are
\begin{align}
e_{2,k}&=\left|h_{2,k}-\Re\left(iM_S/(\pi r_0^2 f)\exp(2i\pi f u)\right)\right|
\nonumber \\
e_{3,k}&=\left|h_{3,k}-\Re\left(iM_S/(4\pi r_0^2 f)e^{-4\pi ir_0 f}\exp(2i\pi f u)\right)\right|\,.
\end{align} 
For the case $u_0=0$ms, $u_{N-1}=100$ms, $f=1$kHz, $r_0=1.25$ms ($\approx $km), and $M_S=0.25$ms ($\approx  M_\odot$), we found:
\begin{equation}
\setlength\arraycolsep{10pt}
\begin{array}{ccc}
N & ||e_{2,k}|| & ||e_{3,k}|| \\
2^9 & 4.0\times 10^{-4} &1.6\times 10^{-3} \\
2^{15} & 1.6\times 10^{-6} &2.7\times 10^{-5} \\
2^{21} & 2.4\times 10^{-8} &4.3\times 10^{-7}
\end{array}
\label{e-normtest}
\end{equation}
where $||e_k||$ is defined to be
\begin{equation}
|| e_k||=\sqrt{\frac{\sum_{k=0}^{N-1}e_k^2}{N}}\,.
\end{equation}
Thus the errors are tending to zero, and $N$ can be chosen so as to attain a desired accuracy.
Note that an error of order machine precision is achieved for special values of the frequency $f=k(N-1)/N/(u_{N-1}-u_0)$ with $k$ an integer; in this case we would have that the cyclic assumption of the discrete Fourier transform would be satisfied, i.e. $u_i=u_{i+N}$.

%%%%%%%%%%%%%%%%%%%%%%%%%%%%%%%%%%%%%%%%%%%%%
\section{Could matter shells explain the GW echo claims?}
\label{s-echoes}

In \cite{Abedi:2016hgu}, the first of the tentative search for echoes, the authors find evidence for the existence of echoes in  the first detection event ~\cite{Abbott:2016blz} {\bf GW150914}. 
They find further comparable evidence for echoes from the events GW151012 ~\cite{Nitz:2018imz} (then referred to as LVT151012) and GW151226 ~\cite{Abbott:2016nmj}.
The references report a number of echo events for GW150914, with the first occurring at about 0.3s after merger; therefore, if caused by a matter shell, the radius would be about 45,000km. The magnitude of the echo was about 0.0992 times that of the original signal. Using 132Hz for the frequency, which is its value when the amplitude was at its maximum at the end of the merger phase, and applying Eq.~(\ref{e-MagEcho}) gives M$_S\simeq$740,000M$_\odot$. Such a mass within a radius of 45,000km would constitute a black hole, so the scenario of an echo caused by a shell can be discounted for GW150914.

Extending their investigations to the first BNS detection {\bf GW170817} ~\cite{Abbott_2017}, the authors of \cite{Abedi:2016hgu} find evidence again of the existence of echoes in the postmerger event ~\cite{Abedi:2018npz}.
The echo was reported to occur at frequency $f_{echo}\simeq 72$ Hz, approximately 1.0 sec after the BNS merger event. The inspiral signal is at 72Hz about 4.0s before merger, so if the reported echo is caused by a matter shell it must have a radius of about $2.5$s$\simeq$ 750,000km. Assuming that the magnitude of the echo signal must be at least $0.01\times$ that of the original signal and applying Eq.~(\ref{e-MagEcho}), it follows that for the echo to be caused by a shell it would have a mass of approximately $10^7M_\odot$. Now, a mass of $10^7M_\odot$ inside a radius of 750,000km would constitute a black hole, so the scenario of an echo caused by a shell can be discounted for GW170817.

The above two examples illustrate the general difficulty of producing a GW echo by means of a matter shell. An echo, in the usual sense, is a repeat of the original signal after a short time delay, which in practice must be at least hundreds of ms, corresponding to a shell radius of at least $\sim$ 20,000km. Now, Eq.~(\ref{e-MagEcho}) shows that the shell mass $M_S=4\pi R r_0^2 f$ which, for expected values of the frequency $f$, will be large -- either implying a black hole and thus not feasible, or requiring an unexpected astrohysical scenario. A much smaller shell radius would avoid these difficulties, but the effect of the shell would be seen as a modification of the original signal, rather than as an echo. Some possible scenarios are presented in the next section.
%%%%%%%%%%%%%%%%%%%%%%%%%%%%%%
\section{GW events and matter shells}
\label{s-GW-other}
%%%%%%%%%%%%%%%%%%%%%%%%%%%%%
\subsection{GW150914 in the presence of a matter shell}
\label{s-GW150914}
%%%%%%%%%%%%%%%%%%%%%%%%%%%%
\begin{figure}
	\includegraphics[width=\columnwidth,angle=0]{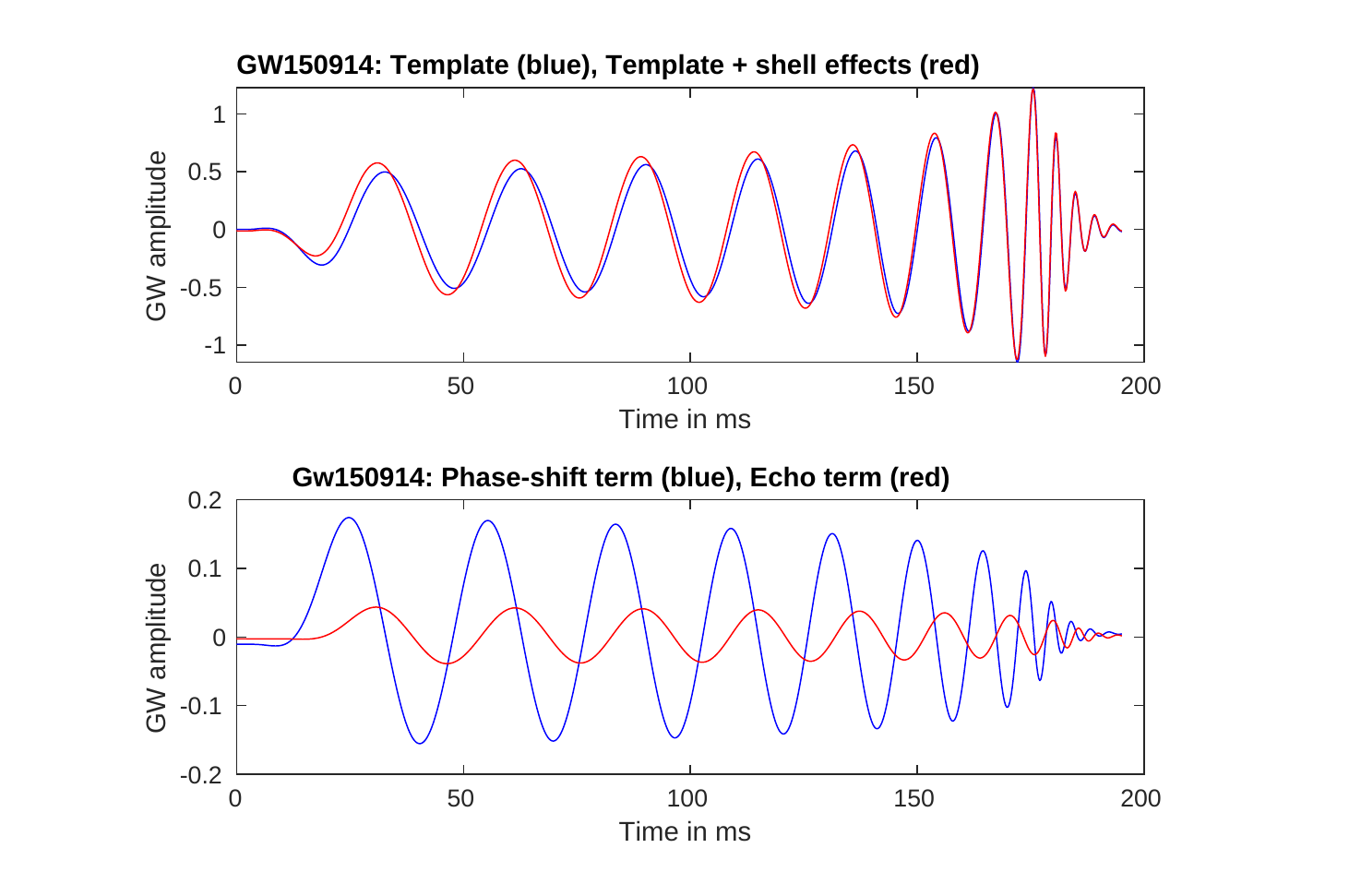} 
	\caption{The effect of a matter shell of radius 3ms (about 900km) and mass 0.3ms (about $60M_\odot$) on the signal of GW150914. The top panel shows the original signal in blue, and the original signal plus modifications due to the shell in red. The lower panel shows the modifications due to the phase-shift term in blue, and due to the echo term in red.}
	\label{fig:BH}
\end{figure}

Although a matter shell could not explain the echoes that might exist in GW150914 data, we now investigate how a hypothetical matter shell might modify the signal from a binary black hole merger. We consider the example case of a shell at radius 3ms (about 900km) and of mass 0.3ms (about $60M_\odot$) and the signal of GW150914 ~\cite{LIGOScientific:2019hgc,NRwaveform,NRwaveform2} (Of course, the astrophysical evidence does not suggest the existence of such a shell). The results are shown in Fig.~\ref{fig:BH}. The top panel shows that there is a small but noticeable modification to the template signal, particularly at early times; this is because the frequency is lower at early times and so the modification effects are larger. The bottom panel shows the contributions of the phase-shift term ($h_2$, blue) and the echo term ($h_3$, red); it is noticeable that, unlike the template signal, these terms decrease in magnitude as the frequency increases with time.

The accuracy of the results presented in Fig.~\ref{fig:BH} is limited since the formalism used in~\cite{Bishop:2019ckc} assumed a weak field GW source, which is not the case for two black holes at merger. In particular, GWs reflected by the shell would be partially absorbed by the black holes, so reducing the magnitude of the echo contribution to the GW signal.

\subsection{Binary Neutron Star (BNS) mergers}
\label{ss-BNS}
%%%%%%%%%%%%%%%%%%%%%%%%
BNS GW events that have been observed include GW170817~\cite{TheLIGOScientific:2017qsa} and GW190425~\cite{Abbott:2020uma}. Of these, GW170817 was at a higher signal to noise ratio and the event was observed post-merger in the electromagnetic spectrum~\cite{Abbott2017c}, indicating that the post-merger object contained a large amount of free matter; we will therefore focus on this event.

The relevant source parameters reported for the event are~\cite{TheLIGOScientific:2016uux,Abbott:2018exr}: total mass $M_1+M_2=2.74M_\odot$, and radii $R_1=10.8$km, $R_2=10.7$km. The reported GW signal increased in amplitude and frequency until 500Hz, at which stage the signal finished; i.e. the signal was observed during the inspiral phase, but ended as the two objects started to merge. The merger probably produced a central remnant of a neutron star or a black hole, in which case GWs in the lowest quasinormal mode would have been produced; however, the frequency of these GWs would be about 1.5 to 3kHz for a neutron star, or about 6kHz for a $2M_\odot$ black hole, which is outside the sensitivity band of the LIGO detectors.

\begin{figure}
	\includegraphics[width=\columnwidth,angle=0]{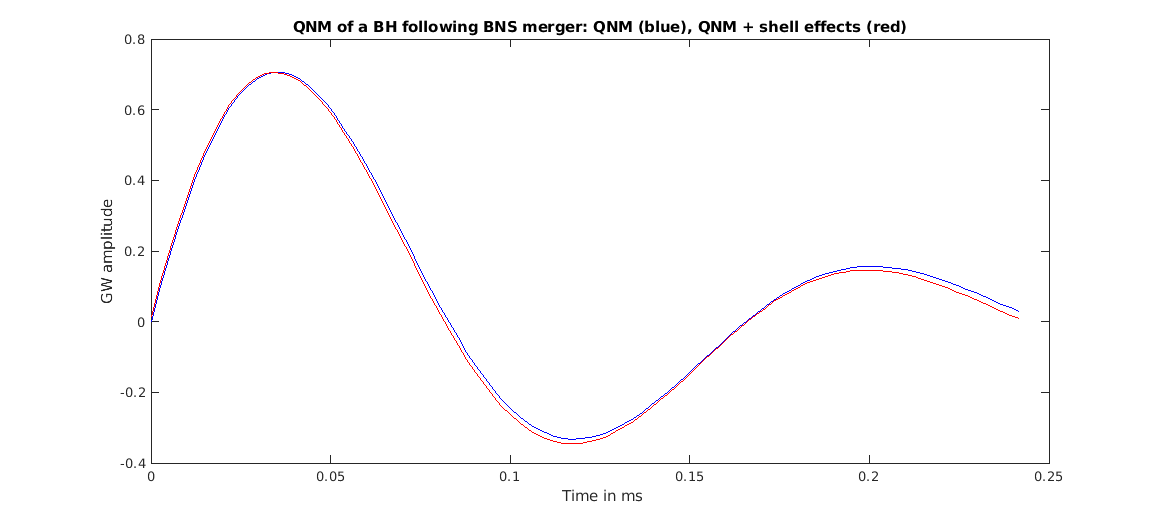} 
	\caption{The effect of a matter shell of radius 25km and mass 0.7$M_\odot$  on a quasinormal mode (QNM) signal of a 2$M_\odot$ remnant of a binary neutron star merger. The original signal is in blue, and the original signal plus modifications due to the shell is in red.}
	\label{fig:BNS}
\end{figure}

In order to estimate the possible effect of matter on GWws emitted from a central remnant, we consider the model of a spherical shell of mass $M_S=0.7M_\odot$ and radius $r_0=25$km around a GW source at either $6$kHz or $2$kHz. We find that the phase-shift term $2iM_S/(r_0^2\nu)$ evaluates to
\begin{equation}
	\frac{iM_S}{r_0^2\pi f}=0.0263i \;\;\mbox{or}\;\; 0.0788i\,,
	\label{e-phtBNS}
\end{equation}
for $6$kHz or $2$kHz, respectively. The echo effect would be $1/4$ of the above values, and the delay would be $0.1667$ms, which is the same as the wave period (at $6$kHz), or less than a wave period (at $2$kHz). Thus, in the future if GW detectors in the kHz band are operational and if a BNS event occurs at high signal to noise ratio, then shell effects would affect the GW signal in a measurable way, although the small delay time means that it would be difficult to disentangle the echo and phase-shift effects from the main signal. The modifications to the quasinormal mode signal are illustrated in Fig.~\ref{fig:BNS}. However, there are a number of caveats that should be noted:
\begin{itemize}
	\item The model in~\cite{Bishop:2019ckc} assumed that the shell is static, but the aftermath of a BNS meger will be highly dynamical.
	\item The hypothesis of a shell forming is not supported by a detailed numerical simulation; indeed, since the system started as an inspiral, the matter outside the remnant should have a ring-like structure. Thus, Eq.~(\ref{e-phtBNS}) may overestimate the matter effect for an observer on the axis of rotation of the system, but may be appropriate for an observer in the equatorial plane.
	\item The comment at the end of Sec.~\ref{s-GW150914} about absorption of GWs by a black hole applies here.
\end{itemize}
So the quantitative values in Eq.~(\ref{e-phtBNS}) should be interpreted as indicative of the order of magnitude of the interaction of GWs with matter, rather than as precise estimates. It should also be noted that if the numerical modeling includes all the matter, and if the simulation run period includes the quasinormal mode ringdown, then shell effects would already be included in the simulation.

%%%%%%%%%%%%%%%%%%%%%%%%%%%%%%%%%%%%%%%%%%%%%%%%%%%%%%%

\subsection{Core collapse Supernovae}
\label{ss-supernova}
%%%%%%%%%%%%%%%%%%%%%%%%%%%%%%%%%%%%%%%%%%%%%%%%%%%%%%%
We next turn our attention to an anticipated candidate of detectable GW waves: core collapse supernovae (CCSNe). 
Whilst binary black hole and BNS mergers are currently the only GW events picked up by LIGO and VIRGO, supernovae are expected to produce, under certain conditions, GW waves detectable by the current generation of interferometers or those on the horizon. 
For now, all detection of supernovae have been confined to electromagnetic detection. 
Photons originate at the outer edge of a star and hence provide only limited information on the interior regions. The detection of GWs which are the result of the aspherical motion of the inner regions will provide a wealth of information on these regions and the mechanism leading to the supernova explosion, where all the four fundamental forces of nature are involved. 

Whilst the central engines and inner regions of CCSNe have yet to be fully understood, there exist several studies of their progenitors and the subsequent evolution and detection ~\cite{Muller:2020ard,Abdikamalov:2020jzn,Woosley02,Woosley:2007as}. 
For stars of mass larger than 8$M_\odot$, evolution normally proceeds through several stages of core burning and then to core collapse once nuclear fusion halts when there are no further burning processes to balance the gravitational attraction. Typically, these cores are iron cores, with the critical mass signalling the onset of core collapse ranging from $1.3M_{\odot}$ to $1.7M_{\odot}$. The core breaks into two during the collapse, with the inner core 
of $0.4M_{\odot}$ to $0.6M_{\odot}$ in sonic contact and collapsing homologously and the outer core collapsing supersonically. 
The inner core reaches supranuclear densities of $\sim 2 \times 10^{14}$g/cm$^{3}$ where the nuclear matter stiffens, resulting in a bounce of the inner core. The resulting shock wave is launched into the collapsing outer core. However, the shock loses energy to dissociation of iron nuclei, stalling  at ${\sim} 150\,\mathrm{km}$ within ${\sim} 10\,\mathrm{ms}$ after formation. 
Many computationally demanding simulations exist 
~\cite{Andresen:2016pdt,Andresen:2018aom,Radice:2018usf} for generation of GWs from CCSNe.

\subsubsection{Model for the GW modifications due to matter around the inner core}
\label{ss-CCSN}
\begin{figure}
	\includegraphics[width=\columnwidth,angle=0]{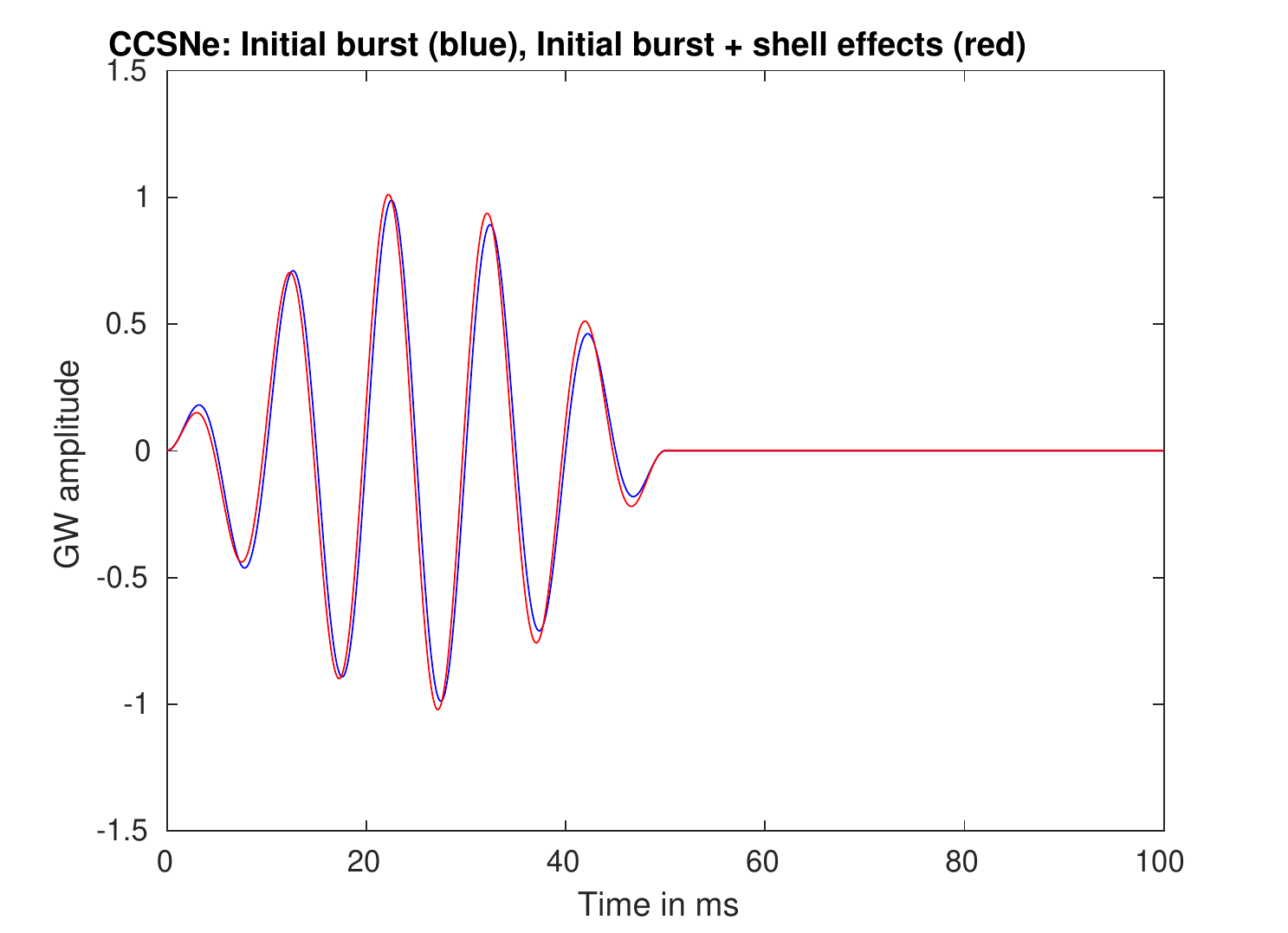}%{FigCCSN.eps} 
	\caption{The effect of a matter shell as specified in the text, on the initial burst (Eq.~(\ref{e-CCSN-burst})) in a CCSNe GW signal. The original signal is in blue, and the modified signal is shown in red.}
	\label{fig:CCSN}
\end{figure}

%%%%%%%%
Ref.~\cite{Radice:2018usf} presents GW waveforms from simulations of CCSNe for various zero age main sequence (ZAMS) masses in the range $9M_\odot,\cdots,60 M_\odot$. The GW signal starts with an initial burst of duration about $50$ms and frequency about $100$Hz, followed by a quiescent period. We model this part of the waveform as
\begin{align}
h_++ih_\times&\sim \sin(0.2\pi  u) \sin(0.02 \pi u)\,{}_2Z_{2,2}\,,\;\; 0\le u\le 50 \nonumber \\
h_++ih_\times&=0\,,\;\; 50\le u\le 100\,.
\label{e-CCSN-burst}
\end{align}
GWs are generated by aspherical motions in the inner core, commencing just after the bounce. The inner core is surrounded by the outer core, treated as a thick matter shell, and we now model its modifications to the GW signal.
The shell has an inner radius $r_{in}$, an outer radius $r_{out}$, and density at $r=r_{in}$ of $\rho_0$ with density fall-off $\rho\propto r^{-1-a}$ with $1/2\le a\le 2$. The effect of the whole shell is obtained by decomposing it into thin shells and then integrating. The result for the echo term is not a simple analytic expression and will be evaluated numerically. However, the phase shift term does give a simple analytic result
\begin{equation}
\int_{r_{in}}^{r_{out}}\frac{iM_s}{r^2 \pi f}dr=i\int_{r_{in}}^{r_{out}}\frac{4\pi r^2 \rho_0 (r_{in})^{1+a}}{r^{1+a}r^2 \pi f}dr
=\frac{4i\rho_0 r_{in}}{af}\left(1-\left(\frac{r_{in}}{r_{out}}\right)^a\right)
\,,
\label{e-CCSN-int}
\end{equation}
where $f$ is the GW frequency. 
Fig.~\ref{fig:CCSN} shows the original signal given by Eq.~(\ref{e-CCSN-burst}) in blue, and the original signal plus shell modifications in red, for the case $r_{in}=0.1$ms ($\approx 30$km), $r_{out}=0.5$ms ($\approx 150$km), $a=1$, and $\rho_0=0.05/$ms${}^{2}$ ($\approx 0.75\times10^{12}$g/cm$^{3}$). These values model: the inner core as a proto-neutron star (PNS) of radius $30$km, and whose oscillations generate the GWs; the shock boundary as having a radius of $150$km; and the density at the inner radius ($\approx 0.75\times10^{12}$g/cm$^{3}$) at a couple of orders below the supranuclear density. For this model, Eq.~(\ref{e-CCSN-int}) evaluates to $0.16i$, and the total mass of the shell to $0.503M_\odot$.

There is some uncertainty in the parameter values that should be used in modeling the matter shell around the inner core, and Eq.~(\ref{e-CCSN-int}) shows how varying the parameters would change the magnitude of the shell effect.
A numerical simulation of a CCSNe entails modeling gravity as well as a number of other physical process, and requires extensive computational resources. It is not feasible to model the whole star using general relativity (GR): approximations to GR are used, and GWs may be estimated using the quadrupole formula~\cite{Radice:2018usf}. Thus corrections to the quadrupolar signal due to shell effects are necessary.

%%%%%%%%%%%%%%%%%%%%%%%%%%%%%%%%%%%%%%%%%%%%%
%Using these we are able to apply our code, taking a typical frequency expected to be around 600Hz, with the shock boundary at $\sim$ 200 km taken as the shell radius. 
%\begin{figure}[h] 
%	\includegraphics[width=0.75\columnwidth,angle=0]{SN.pdf}
%	\caption{The code applied to a model for a supernova.}
%	\label{SNfig}
%\end{figure}
%\FloatBarrier
%%%%%%%%%%%%%%%%%%%%%
%%%%%%%%%%%%%%%%%%%%%%%%%%%%%%%%%%%%%%%%%%%%%
\section{Conclusion}
\label{s-conc}
%%%%%%%%%%%%%%%%%%%%%%%%%%%%%%%%%%%%%%%%%%%%%
There are astrophysical scenarios which can be regarded as comprising a shell of matter around a GW source, and this paper has investigated in what way the GW signal would be affected. The investigation started with GW events for which  echoes have been claimed to exist in the LIGO data, and it was found that such echoes could not be caused by a matter shell. Thus, an unambiguous observation of GW echoes in the future would favour the existence of ECOs.

We investigated the effect of matter shells in three specific example cases. The first was a binary black hole merger analogous to GW150914, surrounded by a hypothetical matter shell at radius $900$km and mass $60M_\odot$. Astrophysically, such a shell is highly unlikely to exist, but this case is useful as it well illustrates some of the features of the shell effects on the waveform. The next case considered was the quasinormal mode signal from the remnant of a binary neutron star merger like GW170917. In this case, it is known that there is a substantial amount of matter around the remnant, although the extent to which the shell model is appropriate is unclear. The final case considered was that of a core collapse supernova. Although GWs from such events have not been observed, they are regarded as potential sources; and here it is clear that the proto neutron star, in which the GWs are generated, is surrounded by shells of matter. Of the three cases, the core collapse supernova is that which yielded the largest shell modifications to the GWs, and for which the predictions are most reliable.

The effects of matter shells are small but measurable if the signal to noise ratio is sufficiently high. As GW observations become more accurate, through both hardware developments and, as time passes, the increasing chance of observing nearby events, these effects will need to be taken into account.

%%%%%%%%%%%%%%%%%%%%%%%%%%%%%%%%%%%%%%%%%%%%%
%\acknowledgments
%This work was supported by the National Research Foundation, South Africa, under grant numbers 118519 and 114815.
%%%%%%%%%%%%%%%%%%%%%%%%%%%%%%%%%%%%%%%%%%%%%
\appendix
%%%%%%%%%%%%%%%%%%%%%%%%%%%%%%%%%%%%%%%%%%%%%
%%%%%%%%%%%%%%%%%%%%%%%%%%%%%%%%%%%%%%%%%%%%%
\section{Matlab scripts}
\label{a-Matlab}
%%%%%%%%%%%%%%%%%%%%%%%%%%%%%%%%%%%%%%%%%%%%%
The Matlab scripts NormTest.m (see Eq.~(\ref{e-normtest})), BH.m used in Sec.~\ref{s-GW150914},  BNS.m used in Sec.~\ref{ss-BNS}, and
CCSN.m used in
Sec.~\ref{ss-CCSN}
are plain text files; the file clean.xlsx is a spreadsheet file containing input data for BH.m. All the files are available as online supplementary material.

\begin{acknowledgements}
This work was supported by the National Research Foundation, South Africa, under grant number 118519.
\end{acknowledgements}
%%%%%%%%%%%%%%%%%%%%%%%%%%%%%%%%%%%%%%%%%%%%
% Authors must disclose all relationships or interests that
% could have direct or potential influence or impart bias on
% the work:
%
%%%%%%%%%%%%%%%%%%%%%%%%%%%%%%%%%%%%%%%%%%%%
\section*{Conflict of interest}
%%%%%%%%%%%%%%%%%%%%%%%%%%%%%%%%%%%%%%%%%%%%
 The authors declare that they have no conflict of interest.
%%%%%%%%%%%%%%%%%%%%%%%%%%%%%%%%%%%%%%%%%%%%
% BibTeX users please use one of
%\bibliographystyle{spbasic}      % basic style, author-year citations
%\bibliographystyle{spmpsci}      % mathematics and physical sciences
\bibliographystyle{spphys}       % APS-like style for physics
%\bibliography{}   % name your BibTeX data base
%%%%%%%%%%%%%%%%%%%%%%%%%%%%

%%%%%%%%%%%%%%%%
%\bibliography{aeireferences,t_1,LitRev9a}
\bibliography{aeireferences,Ref}
%%%%%%%%%%%%%%%%%%%%%%%%%%%%%%%%%%%%%%%%%%%%
\end{document}